\documentclass[aps,prc,twocolumn,superscriptaddress,preprintnumbers,amsmath,amssymb,showkeys,floatfix,nofootinbib,reprint]{revtex4-1}

\usepackage{amsfonts}
\usepackage{amssymb}
\usepackage{graphicx}
\usepackage{dcolumn}
\usepackage{amsmath}
\usepackage{bm}
\usepackage{epsfig}
\usepackage{float}
\usepackage[version=3]{mhchem}
\usepackage{multirow}
\usepackage{url}

\usepackage{color}

\usepackage{longtable}

\usepackage{booktabs}
\usepackage{makecell}

\begin{document}

\title{Isomer depletion via nuclear excitation by inelastic electron scattering}

\author{Ziwen \surname{Li}}%
\affiliation{School of Physics, Nankai University, Tianjin 300071, China}

\author{Jingyan \surname{Zhao}}%
\affiliation{School of Physics, Nankai University, Tianjin 300071, China}

\author{Xuyang \surname{Pu}}%
\affiliation{School of Physics, Nankai University, Tianjin 300071, China}

\author{Yuanbin \surname{Wu}}%
\email{yuanbin@nankai.edu.cn}
\affiliation{School of Physics, Nankai University, Tianjin 300071, China}

\begin{abstract}

Isomer depletion via the process of nuclear excitation by inelastic electron scattering is investigated theoretically. A comprehensive study on low-energy nuclear excitations by inelastic electron scattering is performed to analyze the impact of the nuclear and ion charge, the nuclear transition energy, and the nuclear transition multipolarity on the cross section of the process. We apply the analysis to the case of isomer depletion, in which an excitation from the isomeric state to a nuclear level above the isomeric state can lead to decay to a nuclear level below the isomer itself and hence lead to the release of the energy stored in the isomer. For this purpose, the isomer depletion of $\mathrm{{}^{93m}{Mo}}$, $\mathrm{{}^{152m}{Eu}}$, and $\mathrm{{}^{178m}{Hf}}$, which represent the most important scenarios of isomer depletion, are studied. Our results demonstrate the capability of the process of nuclear excitation by inelastic electron scattering for isomer depletion.

\end{abstract}

\date{\today}

\maketitle


\section{Introduction}

Nuclear isomers are metastable, long-lived excited states of atomic nuclei \cite{Nature(London)399.35,PhysScr.95.044004}. Due to significant differences in spin, nuclear shape, or spin projection along the nuclear symmetry axis, direct transitions of nuclear isomers to lower energy levels are markedly suppressed \cite{Nature(London)399.35,PhysScr.95.044004}. Nuclear isomers have attracted a great deal of attention owing to their importance in the studies in nuclear theory and potential applications in various fields such as nuclear clocks, nuclear gamma-ray lasers, nuclear energy storage, and medical imaging \cite{Nature(London)399.35,PhysScr.95.044004,PhysRevLett.106.162501,RepProPhys.79.076301,NatureRev.Phys.3.238,QuantumSci.Technol.6.034002,PhysRevLett.132.182501,PhysRevLett.133.013201,Nature633.63,PhysRevLett.99.172502,Nature(London)554.216}. Some nuclear isomers exhibit high excitation energy and long half-lives, indicating the possibility for future energy storage. Among the promising candidates, $\mathrm{{}^{178}{Hf}}$ has an isomer with the excitation energy of approximately $2.4$ MeV and a half-life of $31$ years \cite{PhysRevC.68.031302}. Consequently, there is significant interest in developing efficient and controllable methods to release the energy stored in long-lived isomers \cite{Nature(London)399.35,PhysScr.95.044004,PhysRevLett.82.695,PhysRevC.61.054305,PhysRevLett.87.072503,PhysRevC.71.024311,PhysLetB.679,PhysRevLett.99.172502,Nature(London)554.216,YangMRE2025}. The concept of isomer depletion \cite{Nature(London)399.35,Phys.Rep.298,Nat.Phys.1.81,PhysRevLett.82.695,PhysRevLett.99.172502,Nature(London)554.216}, in which the nucleus is excited from the isomeric state to an above-lying level (denoted here as gateway state) which can lead to decay directly to a nuclear level below the isomeric state, is believed to be a promising way to facilitate the depletion of the isomer to release the stored energy on demand.

Driving nuclear transitions via coupling to the atomic shell is one of the most intriguing ways for nuclear excitation. A number of such kind of mechanisms have been proposed for nuclear excitation, including nuclear excitation by electron transition (NEET) \cite{NuclPhysA.539.209,PhysRevLett.85.1831,ProgressofTheoreticalPhysics.49.1574,NuclPhysA.748.3,PhysRevC.76.044611}, nuclear excitation by electron capture (NEEC) \cite{PhysLettB62.393,ContempPhys51.471,PhysRevLett.112.082501,PhysRevLett.40.1695,PhysRevC.59.2462,PhysRevLett.99.172502,JPhysGNuclPartPhys.45.033003,PhysRevLett.128.162501,PhysRevLett.128.212502,ScienceBulletin.67.1526,PhysRevLett.130.112501,EurPhysJ.A.59.281}, nuclear excitation by muon capture (NE$\mu$C) \cite{PhysRevLett.129.142501,PhysRevLett.132.129201,PhysRevLett.132.129202}, and electronic bridge (EB) \cite{PhysRevLett.105.182501,PhysRevLett.121.253002,PhysRevLett.125.032501,PhysRevLett.124.192502,PhysRevC.102.024604,NuclSciTech.32.59,PhysRevLett.133.223001}. NEET/EB involves the bound-bound electronic transition without/with additional photons, while NEEC/NE$\mu$C involves the free-bound electronic/muonic transition. Among them, the resonant process of NEEC, in which a free electron is captured into an atomic shell and the released energy subsequently excites the atomic nucleus, has attracted significant attention recently due to the discussions \cite{PhysRevLett.122.212501,PhysRevLett.127.042501,PhysRevC.108.L031302,Nature(London)594.E1,Nature(London)594.E3,PhysRevLett.128.242502,kbf5-6fcl} on the first reported experimental evidence \cite{Nature(London)554.216} in the $\mathrm{{}^{93m}{Mo}}$ isomer depletion in a beam-based scenario.

In addition to the processes with the direct involvement of bound atomic shells, the process of nuclear excitation by inelastic electron scattering (NEIES) \cite{PhysRev.87.962,PhysRev.96.765,PhysRevLett.124.242501,PhysRevC.106.044604,PhysRevC.106.064604,FrontPhy.1166566,AnnuRevNuclSci.12.1,PhysRevC.59.2462,PhysRevC.110.064621} is also one of the important electron-induced nuclear excitation processes. In the process of NEIES, the nucleus is excited by the electron transition from a higher-energy continuum state to a lower-energy continuum state. Compared to NEET and NEEC with the direct involvement of bound atomic shells, NEIES with the free-free electronic transition has the advantage that the resonant condition with the nuclear transition can be easily fulfilled. It has been experimentally demonstrated that NEIES can be a powerful tool for the isomer production in laser-generated plasmas \cite{PhysRevLett.128.052501,ProcNatlAcadSci.121.e2413221121}.

So far, while more attention has been paid to high-energy electron case, studies of NEIES with low electron energies are still limited \cite{PhysRevC.106.044604}, especially for isomer depletion with low electron energies and low energies of nuclear transitions. In the present work, we focus on the NEIES with the application on isomer depletion. We perform a comprehensive study on low-energy nuclear excitations via the process of NEIES with low electron energies to study the impact of the nuclear and ion charge, the nuclear transition energy, and nuclear transition multipolarity on the cross section of the NEIES process. Then we apply the study to the case of isomer depletion. For this purpose, isomer depletion of $\mathrm{{}^{93m}{Mo}}$, $\mathrm{{}^{152m}{Eu}}$, and $\mathrm{{}^{178m}{Hf}}$, which represent and cover most important scenarios of isomer depletion, are studied.

The paper is organized as follows. In Sec.~\ref{sec:Methods}, we present the theoretical approach for the process of NEIES. Atomic units ($\hbar=m_e=e=1$) are used throughout this section unless otherwise stated. In Sec.~\ref{sec:results}, we present numerical results and discussions. The results of the general case of low-energy nuclear excitations via the process of NEIES with low electron energies, and the influence of various factors including the ion charge, the atomic number, the nuclear transition energy, and nuclear transition multipolarity, on the cross section of NEIES, are presented in Sec.~\ref{sec:results_A}. The results for the isomer depletion of $\mathrm{{}^{93m}{Mo}}$, $\mathrm{{}^{152m}{Eu}}$, and $\mathrm{{}^{178m}{Hf}}$ via NEIES are presented and discussed in Sec.~\ref{sec:results_B}. A brief summary is given in Sec.~\ref{sec:sum}.

\section{Theoretical approach}
\label{sec:Methods}

	We follow the approach of Dirac distorted wave Born approximation in Refs.~\cite{PhysRevC.106.044604,PhysRevC.106.064604,FrontPhy.1166566} to calculate the cross section for the process of NEIES. According to Fermi's golden rule, the differential cross section of nuclear excitation by inelastic electron scattering can be obtained as \cite{PhysRevC.106.044604,PhysRevC.106.064604,FrontPhy.1166566}
	\begin{align}
		\frac{
			\mathrm{d}\sigma
			}{
			\mathrm{d}\Omega_{k_{p_{f}}}
			}
			=\frac{2\pi}{v_{i}}\rho\left(\epsilon_{f}\right)\left|V_{fi}\right|^{2},
	\end{align}
	where $v_{i}$ is the asymptotic incoming speed of the electron, $\rho\left(\epsilon_{f}\right)$ the density of the electron final states, and $\Omega_{k_{p_{f}}}$ the solid angle of the outgoing direction, with $\epsilon_{f}$ and $p_{f}$ being the energy and momentum of the final electron. 
    $V_{fi}$ is the transition matrix element $V_{fi}=\left<f\right|H_{I}\left|i\right>$, where $\left|i\right>$ and $\left|f\right>$ are respectively the initial state and the final state. $H_{I}$ is the interaction Hamiltonian
	\begin{align}
		H_{I}
		&=-\frac{1}{c}\int \left[\vec{j}_{e}\left(\bm{r}\right)
		+\vec{j}_{n}\left(\bm{r}\right)\right]\cdot \vec{A}\left(\bm{r}\right)\mathrm{d}\tau\nonumber\\
		&+\int \frac{\rho_{e}\left(\bm{r}\right)\rho_{n}\left(\bm{r}'\right)}{\left|\bm{r}-\bm{r}^{'}\right|}
		\mathrm{d}\tau\mathrm{d}\tau'.
	\end{align}
    Here $\vec{j}_{e}$ and $\vec{j}_{n}$ are respectively the electron current density operator and the nuclear current density operator. $\rho_{e}$ and $\rho_{n}$ are respectively the charge density of the electron and the nucleus. $\vec{A}\left(\bm{r}\right)$ is the vector potential of the radiation field. The transition matrix element $V_{fi}$ is given as \cite{RevModPhys.28.432,PhysRevC.106.044604,PhysRevC.106.064604,FrontPhy.1166566}
	\begin{align}
		V_{fi}
		&=\sum_{\lambda\mu}\frac{4\pi}{2\lambda+1}\left(-1\right)^{\mu}\nonumber\\
		&\times\left\{\left<\phi_{f}\right|N\left(E\lambda,\mu\right)\left|\phi_{i}\right>\left<I_{f}M_{f}\right|M\left(E\lambda,-\mu\right)\left|I_{i}M_{i}\right>\right.\nonumber\\
		&\left.-\left<\phi_{f}\right|N\left(M\lambda,\mu\right)\left|\phi_{i}\right>\left<I_{f}M_{f}\right|M\left(M\lambda,-\mu\right)\left|I_{i}M_{i}\right>
		\right\}.
	\end{align}
	Here $\left|\phi_{i}\right>$ and $\left|\phi_{f}\right>$ are respectively the initial state and the final state of the electron, and $\left|I_{i}M_{i}\right>$ and $\left|I_{f}M_{f}\right>$ are respectively the initial state and the final state of the nucleus. Here $I_{i,f}$ is the total angular momentum quantum number of the initial state or the final state of the nucleus, and $M_{i,f}$ is the magnetic quantum number of the initial or the final state of the nucleus. $N\left(E\lambda,\mu\right)$ and $N\left(M\lambda,\mu\right)$ are respectively the electric and magnetic multipole transition operators of the electron, with the angular quantum number $\lambda$ and the magnetic quantum number $\mu$. $M\left(E\lambda,\mu\right)$ and $M\left(M\lambda,\mu\right)$ are respectively the electric and magnetic multipole transition operators of the nucleus. $N\left(E\lambda,\mu\right)$, $N\left(M\lambda,\mu\right)$, $M\left(E\lambda,\mu\right)$ and $M\left(M\lambda,\mu\right)$ can be found in Ref.~\cite{RevModPhys.28.432}.

	For the NEIES process, both the electron initial state $\left|\phi_{i}\right>$ and final state $\left|\phi_{f}\right>$ are continuum states. $\left|\phi_{i}\right>$ and $\left|\phi_{f}\right>$ can be expanded into partial wave series \cite{Newton2002}
	\begin{eqnarray}
		\left|\phi_{i}\right>
		&=&\left|\vec{k}_{p_i}\nu_{i}\right>\nonumber\\
		&=&\frac{4\pi}{k_{p_i}}\sqrt{\frac{\epsilon_{i}+m_{e}c^{2}}{2\epsilon_{i}}}\nonumber\\
		&\times& \sum_{\kappa_{i} m_{i}}i^{l_{i}}\Omega^{\dagger}_{\kappa_{i} m_{i}}\left(\hat{k}_{p_i}\right)\chi_{\nu_{i}}e^{id_{\epsilon_{i}\kappa_{i}}}
		\left|\epsilon_{i}\kappa_{i} m_{i}\right>,
	\end{eqnarray}
	and
	\begin{eqnarray}
		\left|\phi_{f}\right>
		&=&\left|\vec{k}_{p_f}\nu_{f}\right> \nonumber\\
		&=&\frac{4\pi}{k_{p_f}}\sqrt{\frac{\epsilon_{f}+m_{e}c^{2}}{2\epsilon_{f}}} \sum_{\kappa_{f} m_{f}}i^{l_{f}}\Omega^{\dagger}_{\kappa_{f} m_{f}}\left(\hat{k}_{p_f}\right)\chi_{\nu_{f}} \nonumber\\
		&\times& e^{-id_{\epsilon_{f}\kappa_{f}}} \left|\epsilon_{f}\kappa_{f} m_{f}\right>.
	\end{eqnarray}
	Here $\vec{k}_{p_i}$ and $\vec{k}_{p_f}$ are respectively the wave vector of the initial electron and the final electron, and $\nu_{i}$ and $\nu_{f}$ spins of the electron. $\kappa_{i,f}$ is the Dirac angular momentum quantum number, which can be determined by the total angular momentum $j_{i,f}$ and the orbital angular momentum $l_{i,f}$. $m_{i,f}$ is the magnetic quantum number of $j_{i,f}$. $d_{\epsilon_{i}\kappa_{i}}$ and $d_{\epsilon_{f}\kappa_{f}}$ are the total phase shifts. $\epsilon_{i}$ and $\epsilon_{f}$ are respectively the energy of the initial electron and the final electron. $\Omega_{\kappa m}$ is the spherical spinor
	\begin{align}
		\Omega_{\kappa m}=\sum_{\nu=\pm 1/2}C\left(l\frac{1}{2}j;m-\nu,\nu,m\right)Y_{l,m-\nu}\chi_{\nu},
	\end{align}
	where the spinors $\chi_{\nu}$ are $\chi_{\frac{1}{2}}=\left(\begin{array}{c}1\\0\end{array}\right)$ and $\chi_{-\frac{1}{2}}=\left(\begin{array}{c}0\\1\end{array}\right)$, and $Y$ is the spherical harmonics. $\left|\epsilon_{i}\kappa_{i} m_{i}\right>$ and $\left|\epsilon_{f}\kappa_{f} m_{f}\right>$ are the relativistic continuum electron wave functions
	\begin{align}
		\left|\epsilon_{i}\kappa_{i}m_{i}\right>=\left(
		\begin{array}{c}
			g_{\epsilon_{i}\kappa_{i}}\left(r\right)\Omega_{\kappa_{i}m_{i}}\\
			if_{\epsilon_{i}\kappa_{i}}\left(r\right)\Omega_{-\kappa_{i}m_{i}}
		\end{array}
		\right),
	\end{align}
	and
	\begin{align}
		\left|\epsilon_{f}\kappa_{f}m_{f}\right>=\left(
		\begin{array}{c}
			g_{\epsilon_{f}\kappa_{f}}\left(r\right)\Omega_{\kappa_{f}m_{f}}\\
			if_{\epsilon_{f}\kappa_{f}}\left(r\right)\Omega_{-\kappa_{f}m_{f}}
		\end{array}
		\right).
	\end{align}
    Here, $g_{\epsilon_{i}\kappa_{i}}$, $f_{\epsilon_{i}\kappa_{i}}$, $g_{\epsilon_{f}\kappa_{f}}$, and $f_{\epsilon_{f}\kappa_{f}}$ are radial wave functions.
    
    The total cross section is then given by
	\begin{eqnarray}
		\sigma_{\mathrm{NEIES}}&=& \frac{2\pi}{v_{i}}\rho\left(\epsilon_{f}\right)
		\frac{1}{2\left(2I_{i}+1\right)}\frac{1}{4\pi} \nonumber \\
        &\times&\int\mathrm{d}\Omega_{k_{p_{i}}}\mathrm{d}\Omega_{k_{p_{f}}}\sum_{M_{i}M_{f}\nu_{i}\nu_{f}}
		\left|V_{fi}\right|^{2}.
	\end{eqnarray}
    
	Performing the average over the initial states, the summation over the final states, and the integration over the electron solid angle, the total cross section can be obtained \cite{PhysRevC.106.044604,PhysRevC.106.064604}
	\begin{eqnarray}
		& &\sigma_{\mathrm{NEIES}}\nonumber\\
		&=&
		\frac{8\pi^{2}}{c^{4}}\frac{k_{p_{f}}}{k_{p_{i}}}
		\frac{\epsilon_{f}+m_{e}c^{2}}{k_{p_{f}}^{2}}\frac{\epsilon_{i}+m_{e}c^{2}}{k_{p_{i}}^{2}}\frac{k^{2\lambda+2}}{\left(2\lambda+1\right)!!^{2}}\nonumber\\
		&\times& B\left(E\lambda,I_{i}\rightarrow I_{f}\right)\sum_{\kappa_{i}\kappa_{f}}\left|M_{\kappa_{f}\kappa_{i}}^{E\lambda}\right|^{2}\left(2j_{i}+1\right)\left(2l_{f}+1\right)\nonumber\\
		&\times& \left(2l_{i}+1\right)\left(2j_{f}+1\right)
		\left\{
		\begin{array}{ccc}
			j_{f}&\lambda&j_{i}\\
			l_{i}&\frac{1}{2}&l_{f}
		\end{array}
		\right\}^{2}
		\left(
		\begin{array}{ccc}
			l_{f}&\lambda&l_{i}\\
			0&0&0
		\end{array}
		\right)^{2}
	\end{eqnarray}
    for electric transition with the multipolarity $\lambda$, and 
    \begin{eqnarray}
		& & \sigma_{\mathrm{NEIES}}\nonumber\\
		&=& \frac{8\pi^{2}}{c^{4}}\frac{k_{p_{f}}}{k_{p_{i}}}
		\frac{\epsilon_{f}+m_{e}c^{2}}{k_{p_{f}}^{2}}\frac{\epsilon_{i}+m_{e}c^{2}}{k_{p_{i}}^{2}}\frac{k^{2\lambda+2}}{\left(2\lambda+1\right)!!^{2}}\nonumber\\
		&\times& B\left(M\lambda,I_{i}\rightarrow I_{f}\right) \sum_{\kappa_{i}\kappa_{f}}\left|M_{\kappa_{f}\kappa_{i}}^{M\lambda}\right|^{2}\nonumber\\
        &\times& \left(2j_{i}+1\right)\left(2j_{f}+1\right) \left(2l_{f}+1\right)\left(2l_{i}^{-}+1\right) \nonumber\\
        &\times& \left\{
		\begin{array}{ccc}
			j_{f}&j_{i}&\lambda\\
			l_{i}^{-}&l_{f}&\frac{1}{2}
		\end{array}
		\right\}^{2}
		\left(
		\begin{array}{ccc}
			l_{f}&\lambda&l_{i}^{-}\\
			0&0&0
		\end{array}
		\right)^{2}
	\end{eqnarray}
    for magnetic transition with the multipolarity $\lambda$. Here, $B\left(E\lambda,I_{i}\rightarrow I_{f}\right)$ and $B\left(M\lambda,I_{i}\rightarrow I_{f}\right)$ are the nuclear reduced transition probabilities, $l_{i}^{-}=2j_{i}-l_{i}$, and $k=\Delta E/c$ with the nuclear transition energy $\Delta E$. $M_{\kappa_{f}\kappa_{i}}^{E\lambda}$ is given by 
	\begin{eqnarray}\label{eq:melec}
		& & M_{\kappa_{f}\kappa_{i}}^{E\lambda}\nonumber\\
		&=& \int h_{\lambda}^{\left(1\right)}\left(kr\right)
		\left[f_{\epsilon_{i}\kappa_{i}}(r)f_{\epsilon_{f}\kappa_{f}}(r)+g_{\epsilon_{i}\kappa_{i}}(r)g_{\epsilon_{f}\kappa_{f}}(r)\right]r^{2}\mathrm{d}r\nonumber\\
		&-& \frac{k}{\lambda}\int h_{\lambda-1}^{\left(1\right)}\left(kr\right) \nonumber\\
        & & \times \left[f_{\epsilon_{i}\kappa_{i}}\left(r\right)f_{\epsilon_{f}\kappa_{f}}\left(r\right)+g_{\epsilon_{i}\kappa_{i}}\left(r\right)g_{\epsilon_{f}\kappa_{f}}\left(r\right)\right]r^{3}\mathrm{d}r\nonumber\\
		&+&
		\frac{k}{\lambda}\int h_{\lambda}^{\left(1\right)}\left(kr\right)\nonumber\\
		& & \times\left[ g_{\epsilon_{f}\kappa_{f}}\left(r\right)f_{\epsilon_{i}\kappa_{i}}\left(r\right)-f_{\epsilon_{f}\kappa_{f}}\left(r\right)g_{\epsilon_{i}\kappa_{i}}\left(r\right)
		\right]r^{3}\mathrm{d}r.
	\end{eqnarray}
    And $M_{\kappa_{f}\kappa_{i}}^{M\lambda}$ is given by
	\begin{eqnarray} \label{eq:mmag}
		M_{\kappa_{f}\kappa_{i}}^{M\lambda}
		&=& \frac{\kappa_{i}+\kappa_{f}}{\lambda}\int h_{\lambda}^{\left(1\right)}\left(kr\right)\left[g_{\epsilon_{f}\kappa_{f}}\left(r\right)f_{\epsilon_{i}\kappa_{i}}\left(r\right)\right.\nonumber\\
		& & \left.+f_{\epsilon_{f}\kappa_{f}}\left(r\right)g_{\epsilon_{i}\kappa_{i}}\left(r\right)\right]r^{2}\mathrm{d}r.
	\end{eqnarray}
    Here $h_{\lambda}^{\left(1\right)}\left(kr\right)$ is the spherical Hankel function of the first kind. We note that there are three terms in Eq.~(\ref{eq:melec}), and for the cases under consideration in the present work, the main contribution is expected to be from the first term while the contribution from the last two terms is small \cite{PhysRevC.106.044604,PhysRevC.106.064604}. Thus, in the present work, we only keep the first term and ignore the second and third terms in Eq.~(\ref{eq:melec}). The calculations in Eqs.~(\ref{eq:melec})-(\ref{eq:mmag}) involve radial wave functions of electrons. In the present work, we adopt the code RADIAL \cite{CPC2019RadialCode} with Dirac-Hartree-Fock-Slater method \cite{PhysRev.137.A27DHFS,CPC1971DHFS} and the Fermi charge distribution to calculate the radial wave functions of electrons.

\section{Numerical results and discussions}
\label{sec:results}

We are now turning to the numerical results and discussions. Firsly, we study the general case of low-energy nuclear excitations via the process of NEIES with low electron energies, and analyze the influence of the ion charge, the atomic number, the nuclear transition energy, and nuclear transition multipolarity, on the cross section of NEIES. Then we apply the study to the case of isomer depletion to study the isomer depletion of $\mathrm{{}^{93m}{Mo}}$, $\mathrm{{}^{152m}{Eu}}$, and $\mathrm{{}^{178m}{Hf}}$ via NEIES. The three candidates $\mathrm{{}^{93m}{Mo}}$, $\mathrm{{}^{152m}{Eu}}$, and $\mathrm{{}^{178m}{Hf}}$ are selected to represent most important scenarios of isomer depletion.

\subsection{General case of low-energy nuclear excitations}
\label{sec:results_A}

\begin{table}[htbp]
\centering
\scalebox{0.932}{
\begin{tabular}{c l l l l }
\hline\hline 
$E_t$ & ~~~~1~$\mathrm{keV}$ & ~~10~$\mathrm{keV}$ & ~~20~$\mathrm{keV}$ & ~~50~$\mathrm{keV}$\\
\hline \multicolumn{5}{c}{$Z=20, A=40$} \\
$E2$ & $9.94\times10^{-6}$ & $9.94\times10^{-1}$ & $3.18\times10^{1}$ & $3.11\times10^{3}$\\
$E3$ & $5.42\times10^{-17}$ & $5.42\times10^{-10}$ & $6.93\times10^{-8}$ & $4.23\times10^{-5}$\\
$E4$ & $1.99\times10^{-28}$ & $1.99\times10^{-19}$ &  $1.02\times10^{-16}$ & $3.89\times10^{-13}$\\
$M1$ & $3.18\times10^{4}$ & $3.18\times10^{7}$ & $2.55\times10^{8}$ & $3.98\times10^{9}$\\
$M2$ & $2.65\times10^{-7}$ & $2.65\times10^{-2}$ & $8.47\times10^{-1} $ & $8.27\times10^{1}$\\
$M3$ & $1.44\times10^{-18}$ & $1.44\times10^{-11}$ & $1.85\times10^{-9}$ & $1.13\times10^{-6}$\\
\hline \multicolumn{5}{c}{$Z=50, A=120$} \\
$E2$ & $4.30\times10^{-5}$ & 4.30 & $1.38\times10^{2}$ & $1.34\times10^{4}$\\
$E3$ & $4.87\times10^{-16}$ & $4.87\times10^{-9}$ & $6.24\times10^{-7}$ & $3.81\times10^{-4}$\\
$E4$ & $3.73\times10^{-27}$ & $3.73\times10^{-18}$ & $1.91\times10^{-15}$ & $7.29\times10^{-12}$\\
$M1$ & $3.18\times10^{4}$ & $3.18\times10^{7}$ & $2.55\times10^{8}$ & $3.98\times10^{9}$\\
$M2$ & $5.50\times10^{-7}$ & $5.50\times10^{-2}$ & 1.76 & $1.72\times10^{2}$\\
$M3$ & $6.24\times10^{-18}$ & $6.24\times10^{-11}$ & $7.99\times10^{-9}$ & $4.87\times10^{-6}$\\
\hline \multicolumn{5}{c}{$Z=100, A=257$} \\
$E2$ & $1.19\times10^{-4}$ & $1.19\times10^{1}$ & $3.80\times10^{2}$ & $3.71\times10^{4}$\\
$E3$ & $2.24\times10^{-15}$ & $2.24\times10^{-8}$ & $2.86\times10^{-6}$ & $1.75\times10^{-3}$ \\
$E4$ & $2.84\times10^{-26}$ & $2.84\times10^{-17}$ & $1.46\times10^{-14}$ & $5.55\times10^{-11}$\\
$M1$ & $3.18\times10^{4}$ & $3.18\times10^{7}$ & $2.55\times10^{8}$ & $3.98\times10^{9}$\\
$M2$ & $9.14\times10^{-7}$ & $9.14\times10^{-2}$ & 2.93 & $2.86\times10^{2}$\\
$M3$ & $1.72\times10^{-17}$ & $1.72\times10^{-10}$ & $2.20\times10^{-8}$ & $1.35\times10^{-5}$\\
\hline\hline
\end{tabular}
}
\caption{The radiative decay rates in units of $\mathrm{s^{-1}}$ for the assumed nuclear transitions for $Z=20$, $Z=50$, and $Z=100$ analysed in Sec.~\ref{sec:results_A}.}
\label{tableI}
\end{table}

\begin{figure*}
\includegraphics[width=0.99\textwidth]{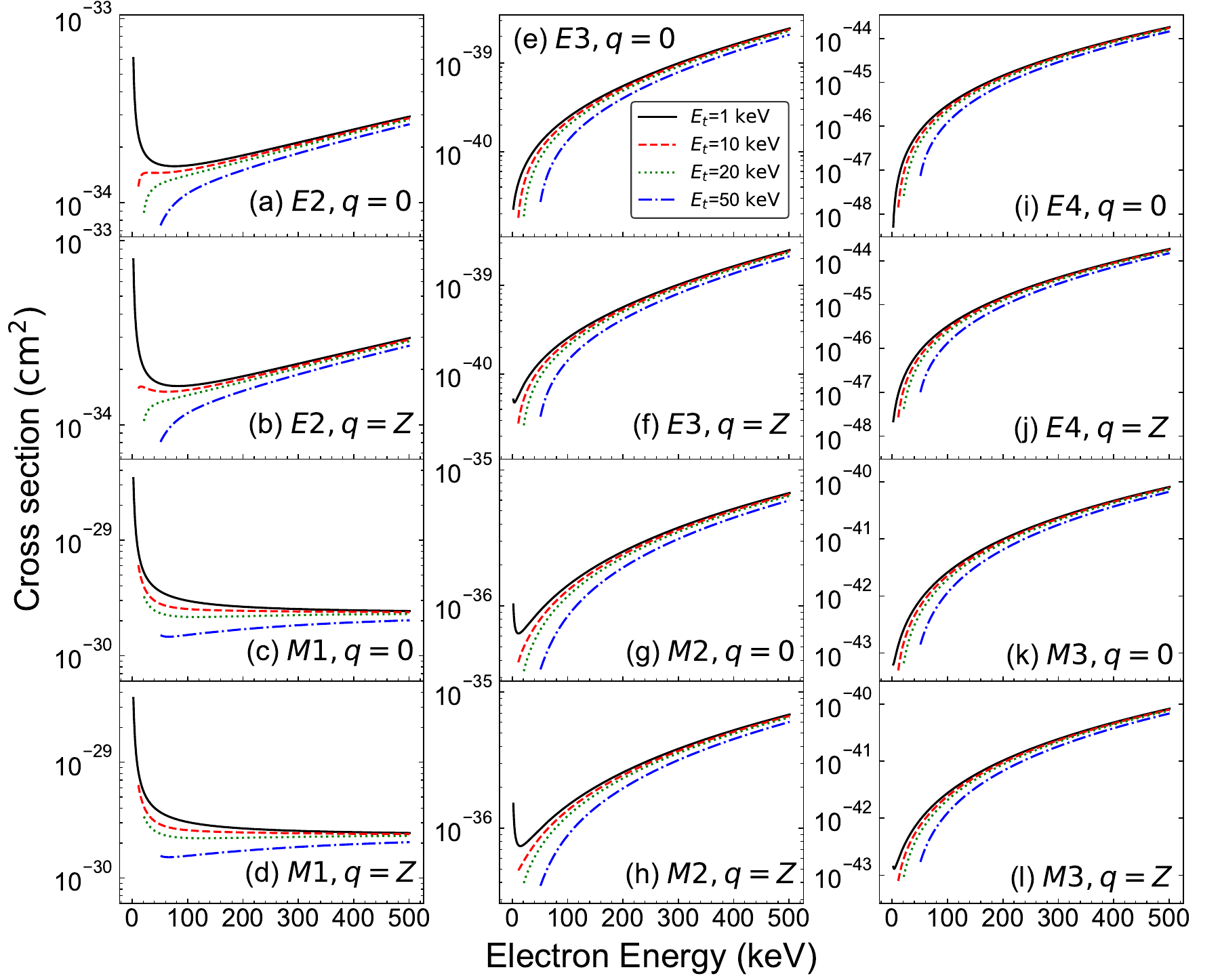}
\caption{ 
The NEIES cross sections for $Z=20$. The results for nuclear transitions of $E2$, $E3$, $E4$, $M1$, $M2$ and $M3$ with selected transition energies $E_t$ are presented. Neutral atoms ($q=0$) and bare nuclei ($q=Z$) are considered.
}
\label{Fig_Z_20_1term}
\end{figure*}

\begin{figure*}
\includegraphics[width=0.99\textwidth]{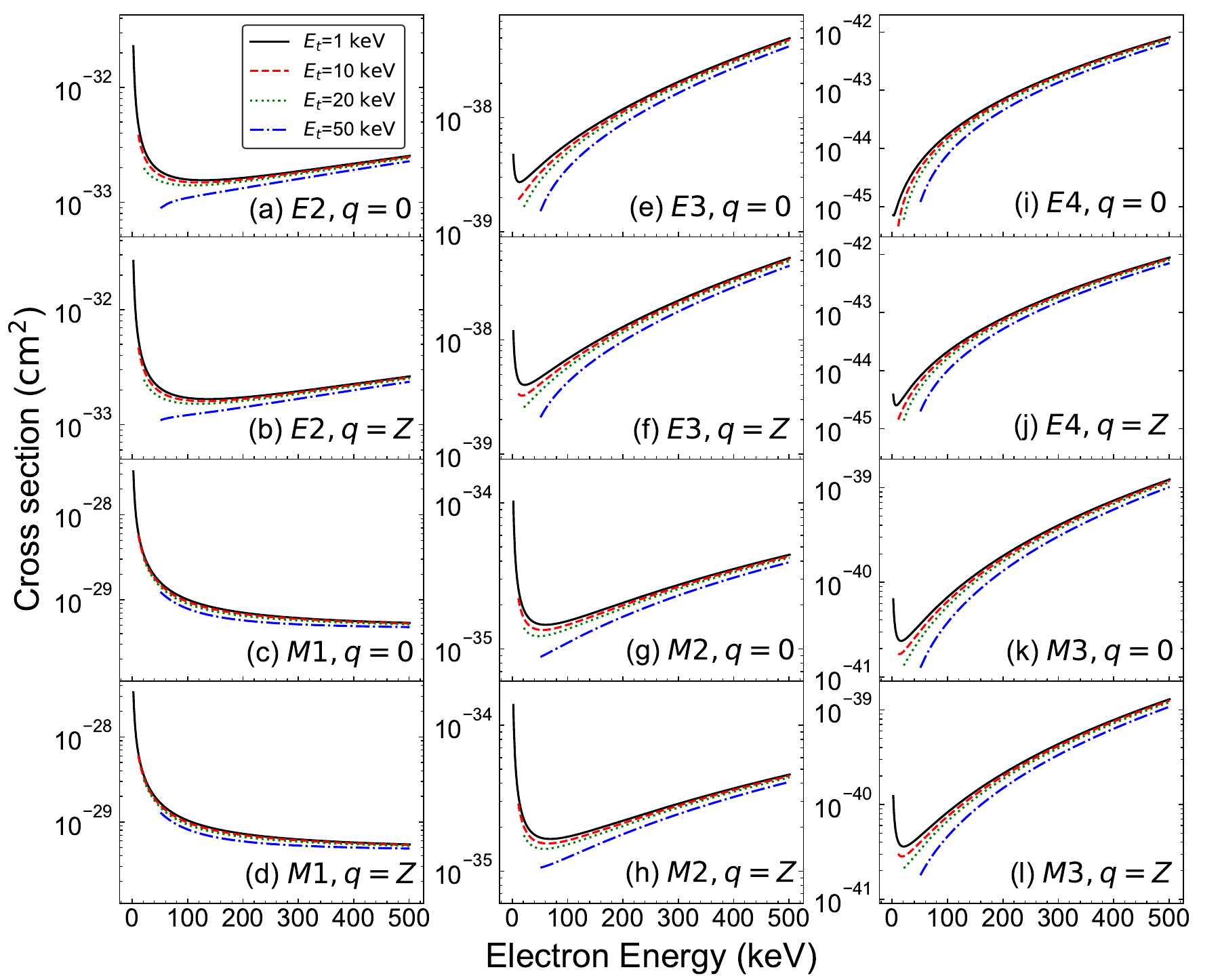}
\caption{ 
The NEIES cross sections for $Z=50$. The results for nuclear transitions of $E2$, $E3$, $E4$, $M1$, $M2$ and $M3$ with selected transition energies $E_t$ are presented. Neutral atoms ($q=0$) and bare nuclei ($q=Z$) are considered. 
}
\label{Fig_Z_50_1term}
\end{figure*}

\begin{figure*}
\includegraphics[width=0.99\textwidth]{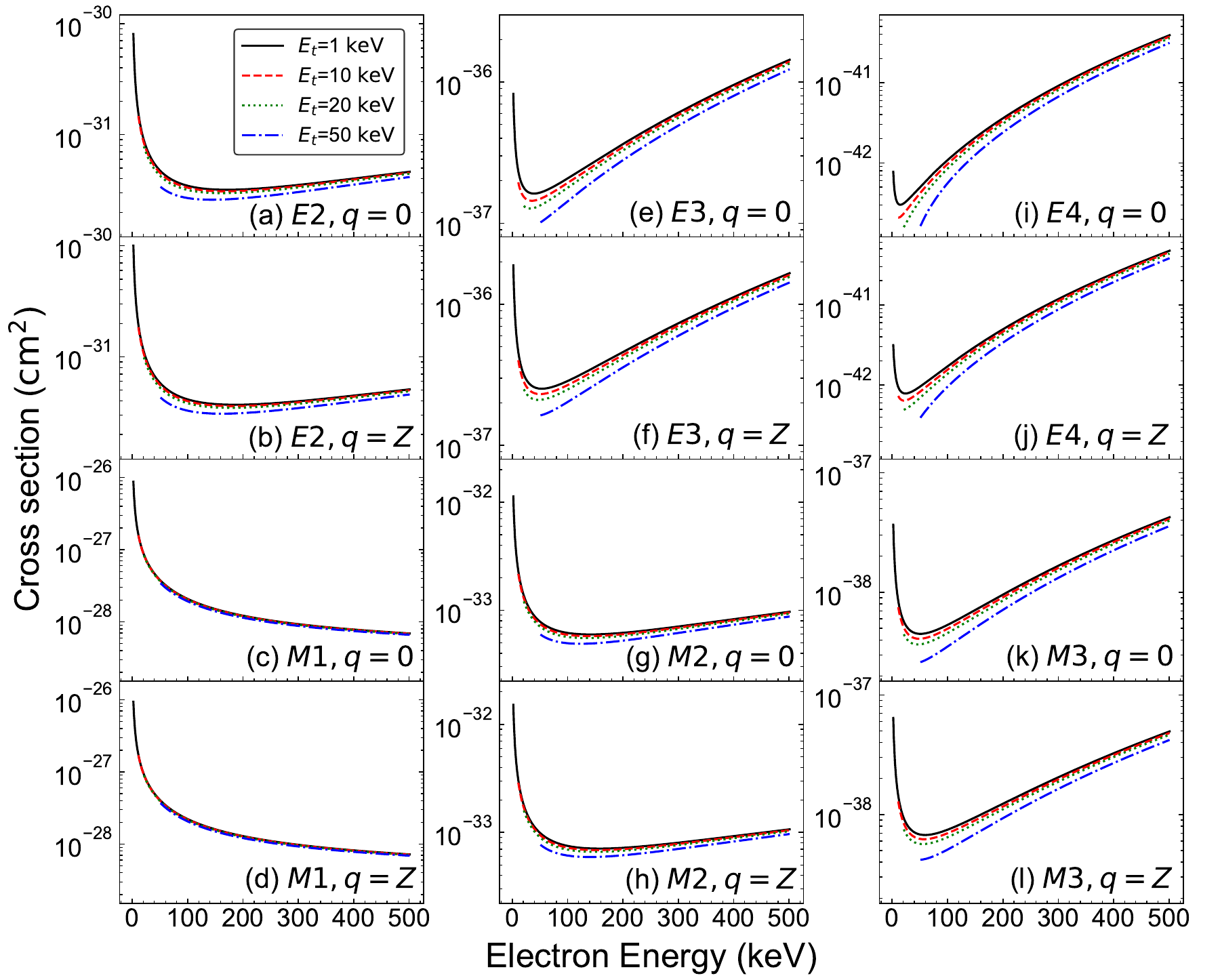}
\caption{ 
The NEIES cross sections for $Z=100$. The results for nuclear transitions of $E2$, $E3$, $E4$, $M1$, $M2$ and $M3$ with selected transition energies $E_t$ are presented. Neutral atoms ($q=0$) and bare nuclei ($q=Z$) are considered.
}
\label{Fig_Z_100_1term}
\end{figure*}

Before restricting our analysis to specific case with an actual nuclide and nuclear transition, we study here the general case of low-energy nuclear excitations to analyze the general features of NEIES with low electron energies. For this purpose, we assume three cases of atomic numbers $Z=20$, $Z=50$, and $Z=100$, which cover the main range of actual atomic numbers. Mass numbers of $40$, $120$, and $257$, are assumed respectively according to the natural abundance of isotopes with the assumed atomic numbers. Nuclear transitions of $E2$, $E3$, $E4$, $M1$, $M2$, and $M3$, which cover most of the important multipolarities of low-energy nuclear transitions, are considered. Nuclear transition energies $E_t=1~\mathrm{keV}$, $10~\mathrm{keV}$, $20~\mathrm{keV}$ and $50~\mathrm{keV}$ are selected. Furthermore, we assume, without loss of generality, nuclear reduced transition probabilities for all cases in this general study to be $1~\mathrm{W.u.}$ (Weisskopf units), i.e., 
\begin{equation}
\begin{aligned}
&B(E2)=B(E3)=B(E4)=1 ~ \mathrm{W.u.},\\
&B(M1)=B(M2)=B(M3)=1 ~ \mathrm{W.u.}.\\
\end{aligned}
\end{equation}
Table~\ref{tableI} shows the radiative decay rates in units of $\mathrm{s^{-1}}$ for the assumed nuclear transitions.

In order to analyze the effect of the ion charge, the two cases of neutral atoms ($q=0$) and bare nuclei ($q=Z$) are assumed here. Electron energies below $500~\mathrm{keV}$ are analyzed here. We have to point out that, the above assumptions do not exactly link to specific and actual nuclide and nuclear transition. Our purpose here is to study general features of NEIES with low electron energies for low-energy nuclear transitions, and analyze the impact of the ion charge, the atomic number, the nuclear transition energy, and nuclear transition multipolarity, on process of NEIES with low electron energies for low-energy nuclear transitions.

Figure~\ref{Fig_Z_20_1term} shows the NEIES cross sections for the case of $Z=20$. It is shown that with a given electron energy, the NEIES cross section decreases with increasing the nuclear transition energy $E_t$, and the differences on the cross sections among different nuclear transition energies decrease with increasing the energy of the incoming electron. This behavior exhibits for all nuclear transition multipolarities $E2$, $E3$, $E4$, $M1$, $M2$ and $M3$ analyzed here. This implies that when the electron energy is much higher than nuclear transition energies, the NEIES process has similar capability to drive nuclear transitions with different transition energies.

We can see in the comparison of the results of neutral atoms and bare nuclei that, the NEIES cross sections for neutral atoms and bare nuclei are on the same order of magnitude throughout the entire electron energy range under consideration, and the NEIES cross section for bare nuclei is slightly larger than the one for neutral atoms. The difference between the cases of neutral atoms and bare nuclei decreases when increasing the energy of the incoming electron. This shows the screening effect of the bound electrons on the NEIES process, and indicates that this screening effect exists, but is rather weak especially for high energy of the incoming electron. The reason may be that only the electron wave functions very close to the nucleus play a dominant role in nuclear excitation, while a large part of the electron cloud of the ion locates beyond this radius. This weak dependence on the ionic states for the process of NEIES is completely deferent from the ones for NEEC and NEET, as in the processes of NEEC and NEET, the cross section depends strongly on the ionic states and the bound electron configurations.

As shown in Fig.~\ref{Fig_Z_20_1term}, the NEIES cross section for $E2$ transition ranges from $10^{-35}$ $\mathrm{cm^2}$ to $10^{-33}$ $\mathrm{cm^2}$, which is about $5$-$6$ orders of magnitude larger than the one for $E3$ transition, and about $10$-$15$ orders of magnitude larger than the one for $E4$ transition. Similarly, Fig.~\ref{Fig_Z_20_1term} also shows that, the NEIES cross section for $M1$ transition spans from $10^{-30}$ $\mathrm{cm^2}$ to $10^{-28}$ $\mathrm{cm^2}$, exhibiting about $5$-$7$ orders of magnitude larger than the one for $M2$ transition and about $10$-$14$ orders of magnitude larger than the one for $M3$ transition. This observation indicates that nuclear transitions with high multipolarities are significantly suppressed. This can be understood as follows. High multipolarities connect to high order transitions in terms of the multipole expansion. For high multipolarities, the selection rules of the NEIES process according to the coupling of the angular momenta require large angular momentum quantum numbers of the contributed partial wave components of the electron. For low electron energies under consideration, contributions involved large angular momentum quantum numbers of the partial wave components of the electron are expected to be suppressed. Furthermore, Fig.~\ref{Fig_Z_20_1term} also clearly shows that, when increasing the angular momentum quantum number $\lambda$, the dependence of the NEIES cross section on the energy of the incoming electron becomes stronger.

It is also shown in Fig.~\ref{Fig_Z_20_1term} that, for $E2$, $M1$, and $M2$ nuclear transitions with enough-low transition energies, the patterns of the curves for the NEIES cross section exhibit an inflection behavior, i.e., the NEIES cross section decreases at first, and subsequently increases ($E2$ and $M2$) or stays almost constant ($M1$), when increasing the energy of the incoming electron. In contrast, for nuclear transition with high-enough transition energies or high multipolarities, the NEIES cross section keeps increasing when increasing the energy of the incoming electron. This implies the strong distorted wave effect for low electron energies and low angular momentum quantum numbers $\lambda$. Furthermore, the results for $E3$ and $M3$ transitions in Fig.~\ref{Fig_Z_20_1term} also show a rather weak inflection behavior on the pattens of the curves of the NEIES cross section for the nuclear transition energy $E_t=1~\mathrm{keV}$ and $q=Z$, while no such a behavior observed for $q=0$. This indicates that the screening effect of the bound electrons weakens the distorted wave effect.

Now we turn to analyze the effect of the atomic number. The NEIES cross sections for $Z=50$ and $Z=100$ with nuclear transitions of $E2$, $E3$, $E4$, $M1$, $M2$ and $M3$ with selected transition energies $E_t$ are presented in Fig.~\ref{Fig_Z_50_1term} and Fig.~\ref{Fig_Z_100_1term}, respectively. Comparison among Figs.~\ref{Fig_Z_20_1term}-\ref{Fig_Z_100_1term} shows that the NEIES cross section in general increases when increasing the atomic number $Z$. For example, for $E2$ transition, the NEIES cross sections are on the order of $10^{-35}$ $\mathrm{cm^2}$ to $10^{-33}$ $\mathrm{cm^2}$ for $Z=20$, $10^{-34}$ $\mathrm{cm^2}$ to $10^{-32}$ $\mathrm{cm^2}$ for $Z=50$, and $10^{-32}$ $\mathrm{cm^2}$ to $10^{-30}$ $\mathrm{cm^2}$ for $Z=100$. This behavior can be easily understood, as the electromagnetic interaction between the electron and the nucleus, which leads to the nuclear excitation, becomes stronger when increasing the atomic number $Z$.

Upon careful examination of the inflection behavior on the pattens of the curves of the NEIES cross section in Figs.~\ref{Fig_Z_20_1term}-\ref{Fig_Z_100_1term} shows that, when increasing the atomic number $Z$, the inflection point occurs at higher energy of the incoming electron. In addition, for nuclear transitions with high transition energies and high multipolarities, the inflection behavior appears and becomes pronounced when increasing $Z$. These can be easily understood, as increasing the atomic number $Z$, the distorted wave effect becomes stronger, and such that the distorted wave effect for higher electron energies and higher angular momentum quantum numbers $\lambda$ could play a role on the NEIES process.

It is shown in Table~\ref{tableI} that the radiative decay rate strongly depends on the nuclear transition energy. When the nuclear transition energy $E_t$ changes from $1~\mathrm{keV}$ to $50~\mathrm{keV}$, one can observe differences about $5$ orders of magnitude for $M1$ transition, about $8$ orders of magnitude for $E2$ transition, about $12$ orders of magnitude for $M3$ transition, and about $15$ orders of magnitude for $E4$ transition, for the radiative decay rate. In contrast, as shown in Figs.~\ref{Fig_Z_20_1term}-\ref{Fig_Z_100_1term}, the dependence on the nuclear transition energy is much weaker for NEIES. For each nucleus, nuclear transition type, and electron energy, the difference between the NEIES cross sections of $E_t = 1~\mathrm{keV}$ and $E_t=50~\mathrm{keV}$ is less than $2$ orders of magnitude, for all cases under consideration. This indicates the difference between the photon-nucleus interaction and the electron-nucleus interaction. Furthermore, comparisons between the radiative decay rates in Table~\ref{tableI} and the NEIES cross sections in Figs.~\ref{Fig_Z_20_1term}-\ref{Fig_Z_100_1term} show that NEIES has a weaker dependence on the nuclear transition multipolarity $\lambda$ than the one for the radiative decay. For example, for the case of $E_t=1~\mathrm{keV}$, the deference between the radiative rates of $E2$ transition and $E4$ transition is about $22$ orders of magnitude, while the deference is about $10$-$15$ orders of magnitude for NEIES. Such behavior indicates that NEIES could play important roles for nuclear transitions with high multipolarities.

\subsection{Isomer depletion via NEIES}
\label{sec:results_B}

\begin{figure}[!htbp]
\includegraphics[width=0.99\linewidth]{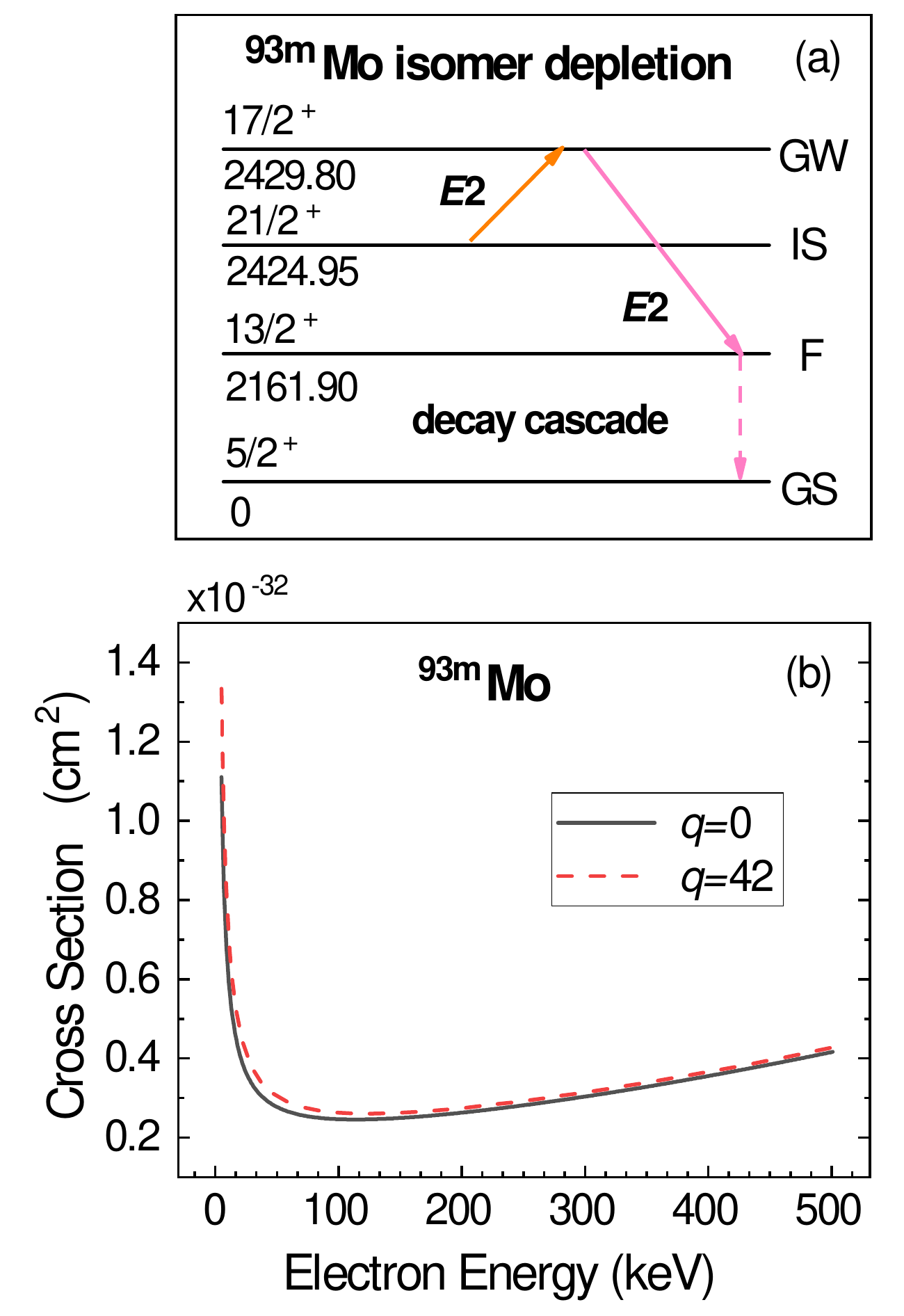}
\caption{ 
 (a) Partial level scheme related to the isomer depletion of $\mathrm{{}^{93m}{Mo}}$. The data of the nuclear levels is taken from Ref.~\cite{nndc}. The energies are presented in units of keV. (b) The NEIES cross sections from the IS to GW of $\mathrm{{}^{93}{Mo}}$ for neutral atoms ($q=0$) and bare nuclei ($q=Z$). 
}
\label{Fig_93Mo}
\end{figure}

\begin{figure}[!htbp]
\includegraphics[width=0.99\linewidth]{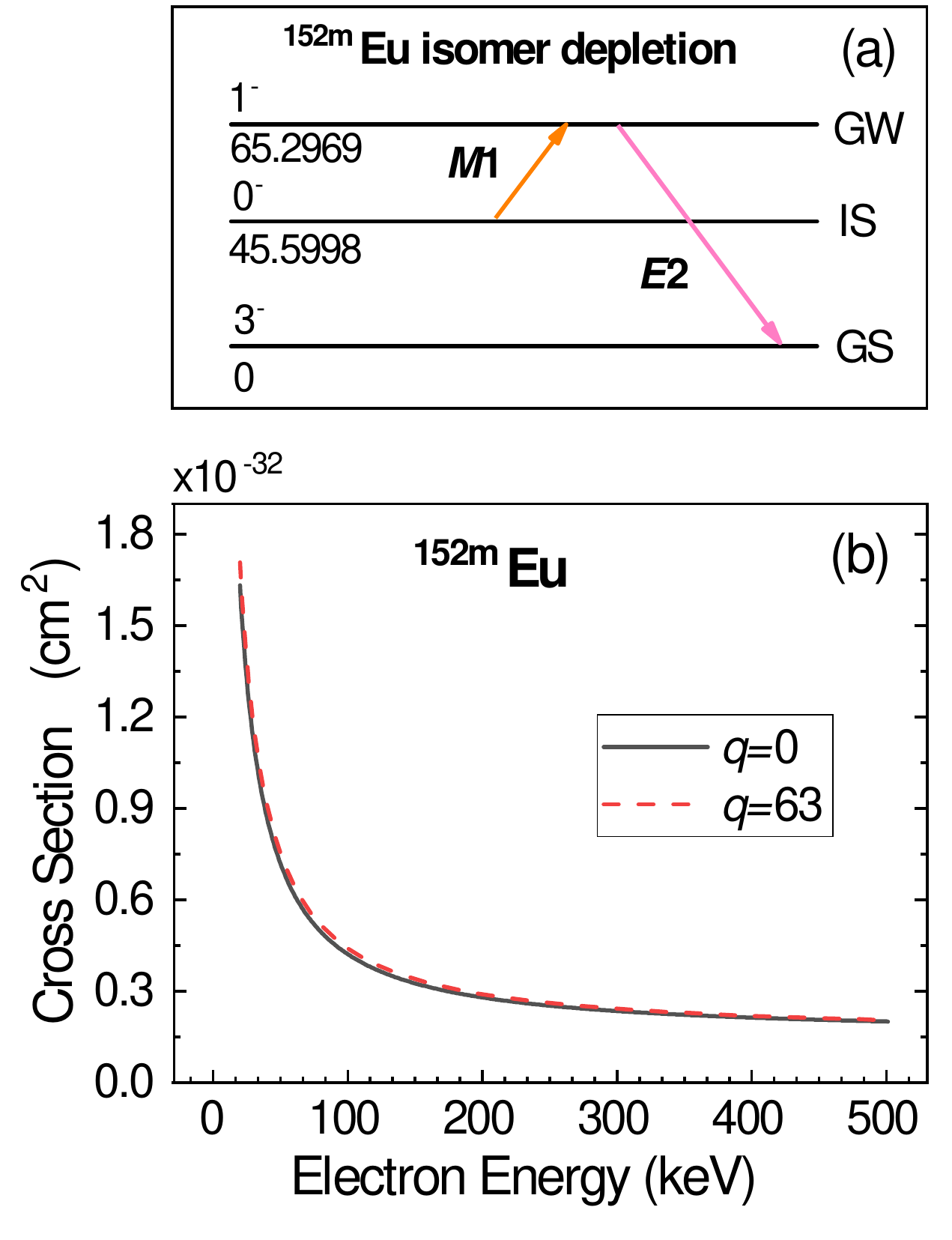}
\caption{(a) Partial level scheme related to the isomer depletion of $\mathrm{{}^{152m}{Eu}}$. The data of the nuclear levels is taken from Ref.~\cite{nndc}. The energies are presented in units of keV. (b) The NEIES cross sections from the IS to GW of $\mathrm{{}^{152}{Eu}}$ for neutral atoms ($q=0$) and bare nuclei ($q=Z$).
}
\label{Fig_152Eu}
\end{figure}

\begin{figure}[!htbp]
\includegraphics[width=0.99\linewidth]{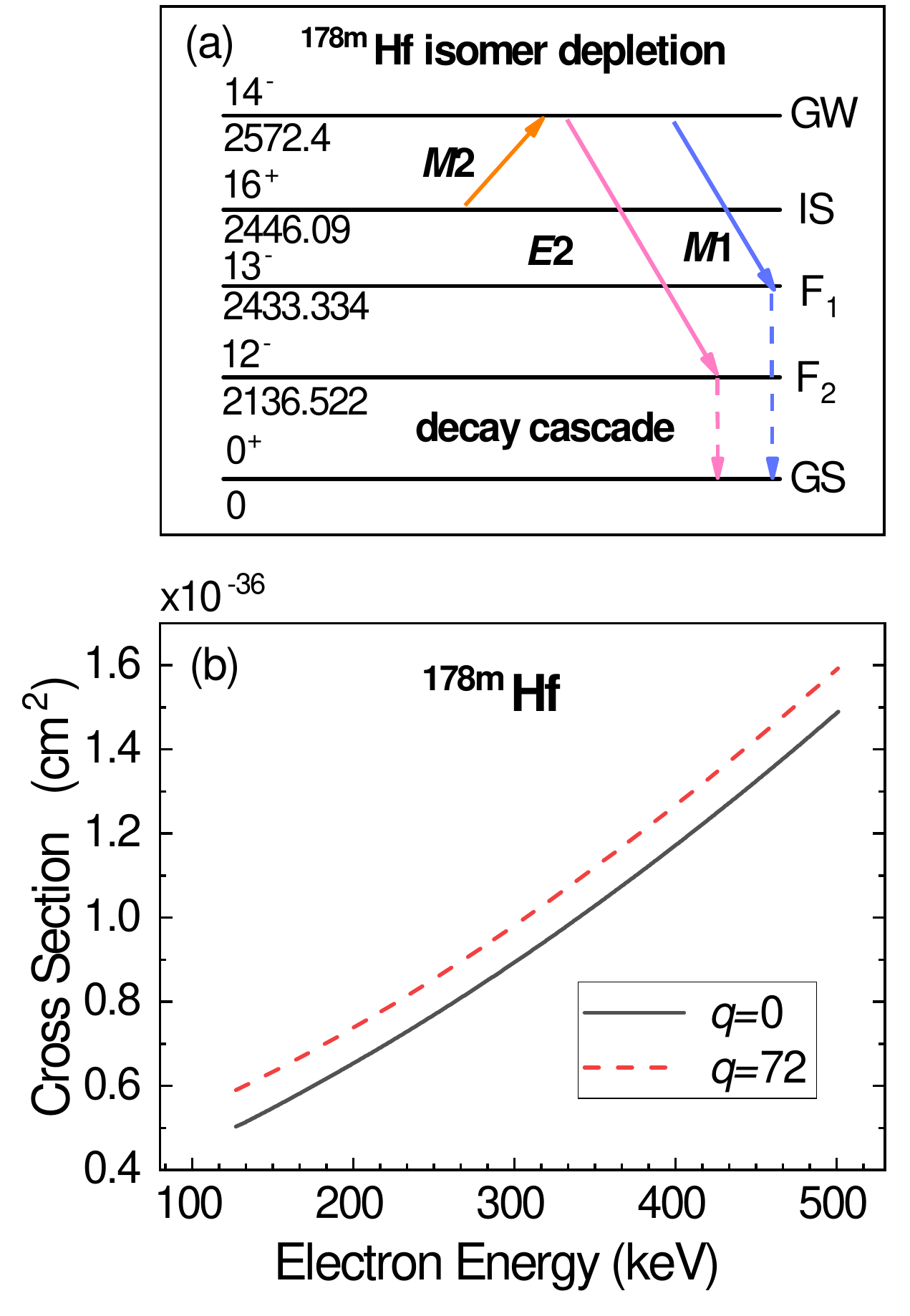}
\caption{(a) Partial level scheme related to the isomer depletion of $\mathrm{{}^{178m}{Hf}}$. The data of the nuclear levels is taken from Ref.~\cite{nndc}. The energies are presented in units of keV. (b) The NEIES cross sections from the IS to GW of $\mathrm{{}^{178}{Hf}}$ for neutral atoms ($q=0$) and bare nuclei ($q=Z$).
}
\label{Fig_178Hf}
\end{figure}

We now turn to the study of isomer depletion via the process of NEIES, i.e., the nucleus at the isomeric state is excited at first from the isomeric state (IS) to an above-lying gateway state (GW) via NEIES, and then subsequently decays to the ground state (GS) or decays to an intermediate state (F) which leads to the decay to the ground state (GS) through a decay cascade. For this purpose, the isomer depletion of $\mathrm{{}^{93m}{Mo}}$, $\mathrm{{}^{152m}{Eu}}$, and $\mathrm{{}^{178m}{Hf}}$ are selected. The isomers $\mathrm{{}^{93m}{Mo}}$ ($6.85~\mathrm{h}$, $2424.95~\mathrm{keV}$),  $\mathrm{{}^{152m}{Eu}}$ ($9.3116~\mathrm{h}$, $45.5998~\mathrm{keV}$) and $\mathrm{{}^{178m}{Hf}}$ ($31~\mathrm{y}$, $2446.09~\mathrm{keV}$) exhibit long lifetimes, rendering them possible candidates for energy storage materials. The nuclear transitions for the isomer depletion of $\mathrm{{}^{93m}{Mo}}$, $\mathrm{{}^{152m}{Eu}}$, and $\mathrm{{}^{178m}{Hf}}$ are the $E2$, $M1$, and $M2$ transitions, respectively, which represent the most important low-energy nuclear transitions and scenarios of isomer depletion. The partial level schemes related to the isomer depletion of $\mathrm{{}^{93m}{Mo}}$, $\mathrm{{}^{152m}{Eu}}$, and $\mathrm{{}^{178m}{Hf}}$ are presented in Figs.~\ref{Fig_93Mo}-\ref{Fig_178Hf}. The data of the nuclear levels in Fig.~\ref{Fig_93Mo}(a), Fig.~\ref{Fig_152Eu}(a), and Fig.~\ref{Fig_178Hf}(a) is taken from the NNDC database \cite{nndc}.

The NEIES cross sections for the nuclear excitation from the isomeric state to the gateway state of $\mathrm{{}^{93}{Mo}}$ are presented in Fig.~\ref{Fig_93Mo}(b). Both neutral atoms ($q=0$) and bare nuclei ($q=Z$) are considered here, and energies of the incoming electron up to $500~\mathrm{keV}$ are considered. Here, we adopt $B(E2)=3.5 ~\mathrm{W.u.}$~\cite{physletb2011Mo93} for the transition from the gateway state to the isomeric state of $\mathrm{{}^{93}{Mo}}$. We can observe from Fig.~\ref{Fig_93Mo} that, the NEIES cross section is on the order of $10^{-33}$ $\mathrm{cm^2}$ to $10^{-32}$ $\mathrm{cm^2}$. And the difference on the NEIES cross sections between the neutral atoms and bare nuclei is small. When increasing the energy of the incident electron, the NEIES cross section first rapidly decreases and then slowly increases, reaching its minimum value at around $80~\mathrm{keV}$ of the energy of the incoming electron.

The NEIES cross sections for the nuclear excitation from the isomeric state to the gateway state of $\mathrm{{}^{152}{Eu}}$ and $\mathrm{{}^{178}{Hf}}$ are presented in Fig.~\ref{Fig_152Eu}(b) and Fig.~\ref{Fig_178Hf}(b), respectively. In our calculation for the case of $\mathrm{{}^{152}{Eu}}$, we adopt the nuclear reduced transition probability as $B(M1)=7.9\times10^{-5}~\mathrm{W.u.}$ for the transition from the gateway state to the isomeric state from the NNDC database \cite{nndc}. For the case of $\mathrm{{}^{178}{Hf}}$, we adopt the nuclear reduced transition probability as $B(M2)=0.0142~\mathrm{W.u.}$ for the transition from the gateway state to the isomeric state from the NNDC database \cite{nndc}. We note that, for the case of $\mathrm{{}^{152}{Eu}}$, the nucleus at the gateway state can decay back to the isomeric state or decay directly from the gateway state to the ground state. For the case of $\mathrm{{}^{178}{Hf}}$, the nucleus at the gateway state has three main decay paths: decay back to the isomeric state and decay to the two intermediate states F$_1$ and F$_2$, as shown in Fig.~\ref{Fig_178Hf}(a). As shown in Fig.~\ref{Fig_152Eu}(b), the NEIES cross section for the nuclear excitation from the isomeric state to the gateway state of $\mathrm{{}^{152}{Eu}}$ is on the order of $10^{-33}$ $\mathrm{cm^2}$ to $10^{-32}$ $\mathrm{cm^2}$, and the cross section keeps decreasing when increasing the energy of the incident electron. For the case of $\mathrm{{}^{178}{Hf}}$ presented in  Fig.~\ref{Fig_178Hf}(b), the NEIES cross section for the nuclear excitation from the isomeric state to the gateway state is on the order of $10^{-37}$ $\mathrm{cm^2}$ to $10^{-36}$ $\mathrm{cm^2}$, and the cross section keeps increasing when increasing the energy of the incident electron. These behaviors are consistent with the general features of NEIES discussed in Sec.~\ref{sec:results_A}.

We note that from the gateway state, the nucleus can decay to the intermediate state(s) or ground state, which refers to isomer depletion, with a branching ratio. Using the nuclear level data from the NNDC database \cite{nndc}, we can obtain the branching ratio. For the case of $\mathrm{{}^{93}{Mo}}$, the branching ratio from the gateway state to the intermediate state is approx. $1$ for both neutral atoms ($q=0$) and bare nuclei ($q=Z$). For the case of $\mathrm{{}^{152}{Eu}}$, the branching ratio from the gateway state to the ground state is approx. $0.1$ for neutral atoms ($q=0$) and approx. $0.2$ for bare nuclei ($q=Z$). For the case of $\mathrm{{}^{178}{Hf}}$, the branching ratio from the gateway state to the intermediate states is approx. $0.5$ for neutral atoms ($q=0$) and approx. $0.9$ for bare nuclei ($q=Z$). Assuming an electron flux of $10^{30}~\mathrm{cm}^{-2}\mathrm{s}^{-1}$ and $10^{10}$ nuclei in the isomeric state in the electron-nucleus interacting region \cite{PhysRevLett.120.052504,PhysRevE.97.063205}, one can expect about $10^8$ events per second for the case of $\mathrm{{}^{93}{Mo}}$, about $10^7$ events per second for the case of $\mathrm{{}^{152}{Eu}}$, and about $10^4$ events per second for the case of $\mathrm{{}^{178}{Hf}}$.

\section{Conclusions}
\label{sec:sum}

In the present work, isomer depletion via the process of nuclear excitation by inelastic electron scattering has been investigated theoretically. We have performed at first a comprehensive study on low-energy nuclear excitations via NEIES with low electron energies. For this purpose, we have assumed three cases of atomic numbers $Z=20$, $Z=50$, and $Z=100$, and nuclear transitions of $E2$, $E3$, $E4$, $M1$, $M2$ and $M3$ with transition energies $E_t=1~\mathrm{keV}$, $10~\mathrm{keV}$, $20~\mathrm{keV}$ and $50~\mathrm{keV}$. Two cases of ion charge states, i.e. neutral atoms ($q=0$) and bare nuclei ($q=Z$), have been considered. Our results clearly show the impact of the nuclear and ion charge, the nuclear transition energy, the energy of the incoming electron, and the nuclear transition multipolarity on the process of NEIES. Our results show that the NEIES process has similar capability to drive nuclear transitions with different transition energies when the energy of the incoming electron is high enough. Our results also show a weak dependence on the ionic states for the process of NEIES as discussed in Refs.~\cite{PhysRevC.106.044604,PhysRevC.106.064604}. The inflection behavior on the patten of the curve of the NEIES cross section and the distorted wave effect on the NEIES process have also been revealed clearly. 

Our results also show that the dependence on the nuclear transition energy and the nuclear transition multipolarity $\lambda$ for NEIES is much weaker than the one for the radiative decay. This indicates that NEIES plays important roles for nuclear transitions with low transition energies or high multipolarities. Then we have applied the analysis to isomer depletion of $\mathrm{{}^{93m}{Mo}}$, $\mathrm{{}^{152m}{Eu}}$, and $\mathrm{{}^{178m}{Hf}}$ via the process of NEIES with low-energy electrons, which represent the most important scenarios of isomer depletion with various nuclear transition energies and transition multipolarities. Our results demonstrate the capability of the process of nuclear excitation by inelastic electron scattering for isomer depletion.

\begin{acknowledgments}
  This work is supported by the National Natural Science Foundation of China (Grant No. 12475122), and by the Fundamental Research Funds for the Central Universities (Grant No. 010-63263118). 
\end{acknowledgments}

\bibliography{papercite}

\end{document}